\documentclass[twocolumn,showpacs,preprintnumbers,amsmath,amssymb]{revtex4}

\usepackage{graphicx}
\usepackage{dcolumn}
\usepackage{bm}
\usepackage{bezier}

\begin{document}

\title{The coexistence of superconductivity and ferromagnetism in nano-scale metallic grains}
\author{Y. Alhassid$^{1}$, K. N. Nesterov$^{1}$ and S. Schmidt$^{2}$}
\affiliation{$^{1}$Center for Theoretical Physics, Sloane Physics
Laboratory, Yale University, New Haven, CT 06520\\
$^{2}$Institute for Theoretical Physics, ETH-Zurich, CH-8093 Zurich, Switzerland}

\date{\today}

\begin{abstract}

A nano-scale metallic grain in which the single-particle dynamics are chaotic is described by the so-called universal Hamiltonian. This Hamiltonian includes a superconducting pairing term and a ferromagnetic exchange term that compete with each other: pairing correlations favor minimal ground-state spin, while the exchange interaction favors maximal spin polarization. Of particular interest is the fluctuation-dominated regime where the bulk pairing gap is comparable to or smaller than the single-particle mean level spacing and the Bardeen-Cooper-Schrieffer theory of superconductivity breaks down. Superconductivity and ferromagnetism can coexist in this regime.  We identify signatures of the competition between superconductivity and ferromagnetism in a number of quantities: ground-state spin, conductance fluctuations when the grain is weakly coupled to external leads and the thermodynamic properties of the grain, such as  heat capacity and spin susceptibility.
\end{abstract}
\maketitle

\newcommand{\bS}{{\hat{\bf S}}}
\newcommand{\hP}{{\hat{P}}}
\newcommand{\Tr}{{\mathrm{Tr}}}
\newcommand{\Th}{{\mathrm{Th}}}
\newcommand{\HBCS}{\hat{H}_{\mathrm{BCS}}}
\newcommand{\Auxfield}{{\widetilde{\Delta}}}
\newcommand{\Dstatic}{{|\Delta_{0}|}}
\newcommand{\Dstaticarg}{{\Delta_0}}
\newcommand{\Hstatic}{{\hat{H}_\Dstaticarg}}

\setcounter{topnumber}{1} 

\section{Introduction}
\label{sec:Introduction}

Advances in nanofabrication techniques in the mid-1990s have made it possible to study the properties of individual ultra-small metallic grains with linear dimension of several nanometers (see Ref.~\cite{vonDelft2001} for a review). Discrete energy levels of such nano-scale grains were measured by single-electron tunneling spectroscopy~\cite{Ralph1995, Black1996, Ralph1997}. Grains made of various materials have been studied, probing different regimes of electron-electron interaction and spin-orbit scattering. Recent experimental techniques allow for better control of the shape and size of the grain~\cite{Kuemmeth2008}.

Superconductivity in bulk metals was explained by the Bardeen-Cooper-Schrieffer (BCS) mean-field theory~\cite{BCS}.  Superconducting correlations in a finite-size system are characterized by the ratio of the bulk pairing gap $\Delta$ to the mean single-particle level spacing $\delta$. BCS theory is valid in the bulk limit $\Delta/\delta \gg 1$ but breaks down in the fluctuation-dominated regime $\Delta/\delta \lesssim 1$. In the first spectroscopy experiments on metallic grains~\cite{Ralph1995, Black1996, Ralph1997}, both regimes were accessed by studying aluminum grains of different sizes. A pairing gap was observed in the excitation spectra of the larger grains with an even number of electrons. However, in the smallest grains, such a pairing gap could not be resolved. These experiments generated much interest in superconductivity in finite-size systems and, in particular, in the fluctuation-dominated regime. It was proposed that pairing correlations in the fluctuation-dominated regime can be observed through the number-parity dependence of thermodynamic observables such as the heat capacity and spin susceptibility~\cite{DiLorenzo2000,Falci2000,Schechter2001,Falci2002}. For example, the spin susceptibility of an odd grain  exhibits a re-entrant behavior (i.e., a local minimum) as the temperature decreases~\cite{DiLorenzo2000}.

The linear size of the ultra-small grains studied in spectroscopy experiments is typically smaller than the mean free path of the electrons inside the grain, and the single-particle dynamics are dominated by scattering from the boundaries of the grain. In grains with irregular boundaries, the single-particle dynamics are chaotic. A characteristic energy scale in such grains is the Thouless energy $E_\Th$, which is determined by the time required for an electron at the Fermi energy to cross the grain. Energy spectra of chaotic grains fluctuate from sample to sample. In an energy window of width $E_\Th$ around the Fermi energy, the statistical properties of the single-particle energies and wave functions are described by random-matrix theory (RMT)~\cite{Mehta1991,Beenakker1997,Guhr1998,Alhassid2000}.

When the dimensionless Thouless conductance $g_\Th= E_\Th/\delta$ is large, a metallic grain with chaotic single-particle dynamics is described by the so-called universal Hamiltonian~\cite{Kurland2000, Aleiner2002}. The single-particle part of this Hamiltonian follows RMT, while the interaction is composed of three universal terms: a classical charging energy term that depends on the capacitance of the grain, a BCS-like pairing term (which is suppressed in the presence of an external orbital magnetic field), and a spin exchange term (in the absence of spin-orbit scattering).

The single-particle part of the universal Hamiltonian and the pairing term favor minimization of the total spin, while the ferromagnetic exchange term tends to polarize the grain. In the bulk limit $\Delta/\delta \gg 1$, the ground state for an even particle number is either completely paired or completely polarized,  depending on the ratio $J_s/\delta$, where $J_s$ is the spin exchange coupling constant. There is a first-order phase transition between the superconducting and ferromagnetic phases, whose signature is the macroscopic onset of magnetization due to the Stoner instability. In a finite system, there is strictly speaking no phase transition, and the ground-state spin of the grain increases stepwise as a function of $J_s/\delta$~\cite{Kurland2000,Schmidt2007}. In particular, in the fluctuation-dominated regime $\Delta/\delta \lesssim 1$, there is a region in the ground-state phase diagram where the electrons in the grain are partly paired and partly polarized, i.e., superconducting and ferromagnetic correlations coexist in the ground-state wave function~\cite{Ying2006,Schmidt2007}.

Here we discuss various signatures of the competition between pairing and exchange interactions in ultra-small metallic grains that are described by the universal Hamiltonian in the absence of orbital magnetic field and spin-orbit scattering~\cite{Schmidt2007,Schmidt2008,VanHoucke2010,Nesterov2012}.

The outline of this paper is as follows. In Sec.~\ref{sec:Model} we discuss the universal Hamiltonian and its general solution. In Sec.~\ref{sec:Phase_Diagram} we describe the ground-state phase diagram of the universal Hamiltonian with an equally spaced single-particle spectrum~\cite{Schmidt2007} and discuss mesoscopic fluctuations of the ground-state spin. In Sec.~\ref{sec:Transport} we discuss signatures of the competition between superconductivity and ferromagnetism in the mesoscopic fluctuations of the linear conductance of a grain that is weakly coupled to external leads~\cite{Schmidt2008}. In Sec.~\ref{sec:Thermodynamics}, we discuss  the signatures of this competition in thermodynamic observables of a nano-size grain and the mesoscopic fluctuations of these observables~\cite{VanHoucke2010,Nesterov2012}. We conclude in Sec.~\ref{sec:Conclusions}.

\section{The universal Hamiltonian}\label{sec:Model}

In the absence of spin-orbit scattering and orbital magnetic field, the universal Hamiltonian for a fixed number of electrons has the form~\cite{Kurland2000, Aleiner2002}
\begin{equation}\label{universal_hamiltonian}
\hat{H} = \sum_{i, \sigma=\uparrow,\downarrow} \epsilon_i c^\dagger_{i\sigma}
c_{i\sigma} - G \hP^\dagger\hP - J_s \bS^2\,,
\end{equation}
where
\begin{equation}\label{pairing operators P+ and P}
\hP^\dagger = \sum_i c^\dagger_{i\uparrow}c^\dagger_{i\downarrow}\quad\text{and}\quad \hP =
\sum_i c_{i\downarrow} c_{i\uparrow}
\end{equation}
are the pair creation and annihilation operators, and $\hat{\bf{S}}$ is the total spin of the electrons in the grain. The single-particle energies $\epsilon_i$ follow the statistics of the Gaussian orthogonal ensemble (GOE) of RMT~\cite{Mehta1991}.  When the particle number $N$ is not fixed (e.g., in the presence of a varying gate voltage), an additional charging energy term $e^2\hat{N}^2/2C$ (where $C$ is the capacitance of the grain) must be included in (\ref{universal_hamiltonian}).

The randomness of the single-particle eigenstates $|i\rangle$ induce fluctuations in the electron-electron interaction matrix elements $v_{ij;kl}$, and the latter form an induced two-body ensemble~\cite{Alhassid2005}. The universal terms in (\ref{universal_hamiltonian}) follow from the average part of this induced random interaction.  This can be derived from the requirement that the average interaction $\overline{v}_{ij;kl}$ be invariant under a change of the single-particle basis (the GOE, which describes the one-body part of the Hamiltonian, is invariant under an orthogonal transformation of the single-particle basis).  There are three orthogonal invariants: $\delta_{ik}\delta_{jl}$, $ \delta_{ij}\delta_{kl}$ and $\delta_{il}\delta_{jk}$. We then have
\begin{equation}
\overline{v}_{ij;kl} = v_0 \delta_{ik}\delta_{jl} -G \delta_{ij}\delta_{kl} + J_s\delta_{il}\delta_{jk} \;,
\end{equation}
where $v_0, G$ and $J_s$ are constants. When expressed in its second-quantized form, this average interaction leads to the interaction terms of the universal Hamiltonian (\ref{universal_hamiltonian}) (where we have omitted the charging energy term related to $v_0$).  The fluctuating part of the interaction is suppressed by $1/g_\Th$~\cite{Kurland2000, Aleiner2002} and can be ignored in the limit $g_\Th \gg 1$.

In practical calculations, the computational effort can be reduced by truncating the band width of the single-particle space. The pairing coupling constant $G$ is then renormalized according to \cite{Alhassid2007, Berger1998}
\begin{equation}
   \frac{G}{\delta} =
    \frac{1}{\mathrm{arcsinh}\left(\frac{N_{\text{sp}}/2}
    {\Delta/\delta}\right)}\;,
\end{equation}
where $N_{\text{sp}}$ is the number of single-particle orbitals. The  exchange coupling constant $J_s/\delta$ (expressed in units of the mean level spacing), is a material-dependent parameter. Its values for different elements were tabulated in Ref.~\cite{Gorokhov2004}.

\subsection{Solution to the universal Hamiltonian}\label{solution}

The universal Hamiltonian is an integrable model, which can be solved by generalizing Richardson's solution of the pairing problem~\cite{Richardson1963, Richardson1967} to include the exchange interaction~\cite{Alhassid2003,Tureci2006,Schmidt2007}. Below we discuss this solution and the quantum numbers used to classify the many-body eigenstates.

We decompose the set of single-particle orbitals into a set ${\cal B}$ of singly occupied orbitals and the complementary set ${\cal U}$ of empty and doubly occupied orbitals. The pairing interaction scatters only time-reversed pairs of electrons from doubly occupied orbitals to empty orbitals. We can thus solve a reduced BCS-like Hamiltonian within the set ${\cal U}$
\begin{equation}\label{BCS_reduced}
 \hat{H}_{\cal U} = \sum_{i\in{\cal U}, \sigma} \epsilon_i c^\dagger_{i\sigma}
c_{i\sigma} - G \hP^\dagger\hP \;.
\end{equation}
We denote the eignestates of (\ref{BCS_reduced}) by $|\zeta\rangle_{\cal U}$.  The singly occupied orbitals in $\cal B$ are not affected by pair scattering and are referred to as ``blocked'' orbitals. They thus form a set of good quantum numbers, and we can solve for a reduced spin exchange Hamiltonian within the set ${\cal B}$
\begin{equation}\label{exchange_reduced}
\hat{H}_{\cal B} = \sum_{i\in{\cal B}, \sigma} \epsilon_i c^\dagger_{i\sigma}
c_{i\sigma} - J_s \bS^2 \;.
\end{equation}
The eigenstates of (\ref{exchange_reduced}) are found by coupling the spins $1/2$ of the electrons in the set $\cal B$ to total spin $S$ and spin projection $S_z$.  We denote the corresponding eigenstates by $|\gamma S S_z\rangle_{\cal B}$,  where $\gamma$ are quantum number used to distinguish between states with the same $S$ and $S_z$~\cite{Alhassid2003,Tureci2006}.
The orbitals in ${\cal U}$ have spin zero and do not contribute to the total spin. The Hamiltonians in (\ref{BCS_reduced}) and (\ref{exchange_reduced}) commute, and the eigenstates of the universal Hamiltonian (\ref{universal_hamiltonian}) are a direct product $|\zeta\rangle_{\cal U} \otimes |\gamma S S_z\rangle_{\cal B}$ with energies $E = E_{\cal U} + E_{\cal B}$.

The eigenstates of the reduced BCS model (\ref{BCS_reduced}) can be found by using Richardson's solution~\cite{Richardson1963, Richardson1967}. A solution with $m$ pairs is expressed in terms of $m$ Richardson parameters $R_\mu$ that satisfy the following $m$ coupled non-linear equations ($\mu=1,\ldots m$)
\begin{equation}
 \frac 1G + 2\sum_{\nu=1 \atop \nu\ne \mu}^m \frac{1}{R_\nu-R_\mu} = \sum_{i \in {\cal U}}\frac{1}{2\epsilon_i-R_\mu}\,.
\end{equation}
The eigenenergy of the reduced BCS-like Hamiltonian (\ref{BCS_reduced}) is the sum of these parameters
\begin{equation}
 E_{\cal U} = \sum_{\mu=1}^m R_\mu \;,
\end{equation}
and the corresponding eigenstate can be written as
\begin{equation}
 |\zeta \rangle_{\cal U} \propto \prod_{\mu=1}^m \left(\sum_{i \in {\cal U}} \frac{1}{2\epsilon_i - R_\mu} c^\dagger_{i\uparrow} c^\dagger_{i\downarrow}\right)|0\rangle\;.
\end{equation}
In the limit $G \to 0$, different configurations of doubly occupied orbitals are not mixed, and for any Richardson parameter $R_\mu$ there is a doubly occupied orbital $k$ such that $R_\mu \to 2\epsilon_k$. There is a one-to-one correspondence between this non-interacting solution and a solution for a finite pairing coupling constant $G$. Thus the quantum numbers $\zeta$ of a pairing solution can be chosen to be the set of doubly occupied orbitals in the corresponding non-interacting solution. In fact, the values of $R_\mu$ can be found numerically by evolving the initial non-interacting values as a function of increasing $G$.

For a given set ${\cal B}$ with $b=N-2m$ blocked orbitals, the number of different eigenstates $|\gamma S S_z\rangle_{\cal B}$ is $2^b$. Their energies
 \begin{equation}
 E_{\cal B} = \sum_{i \in {\cal B}} \epsilon_i - J_s S(S+1)
\end{equation}
are independent of $\gamma$. The number of such states with given $S$ and $S_z$ is~\cite{Alhassid2003}
\begin{equation}
d_b(S) = \left(\begin{array}{c}
b \\
S+b/2
\end{array}\right) - \left(\begin{array}{c}
b \\
S+1+b/2
\end{array}\right) \;.
\end{equation}

\section{Ground-state spin}\label{sec:Phase_Diagram}

In this section we present results for the ground-state spin of the universal Hamiltonian~(\ref{universal_hamiltonian}) as a function of the pairing gap $\Delta/\delta$ and exchange interaction strength $J_s/\delta$ (both measured in units of $\delta$). We find the lowest energy $E(S)$ for a given spin $S$ using the method discussed in Sec.~\ref{solution} and then minimize $E(S)$ with respect to $S$. The ground-state spin of the grain is determined by the competition between various terms in the universal Hamiltonian. The one-body part (kinetic energy plus one-body confining potential) and the pairing interaction favor minimal spin, while the exchange interaction favors a maximally polarized state.

\subsection{Equidistant single-particle spectrum}

\begin{figure}[t]
\centerline{\includegraphics[width=3in]{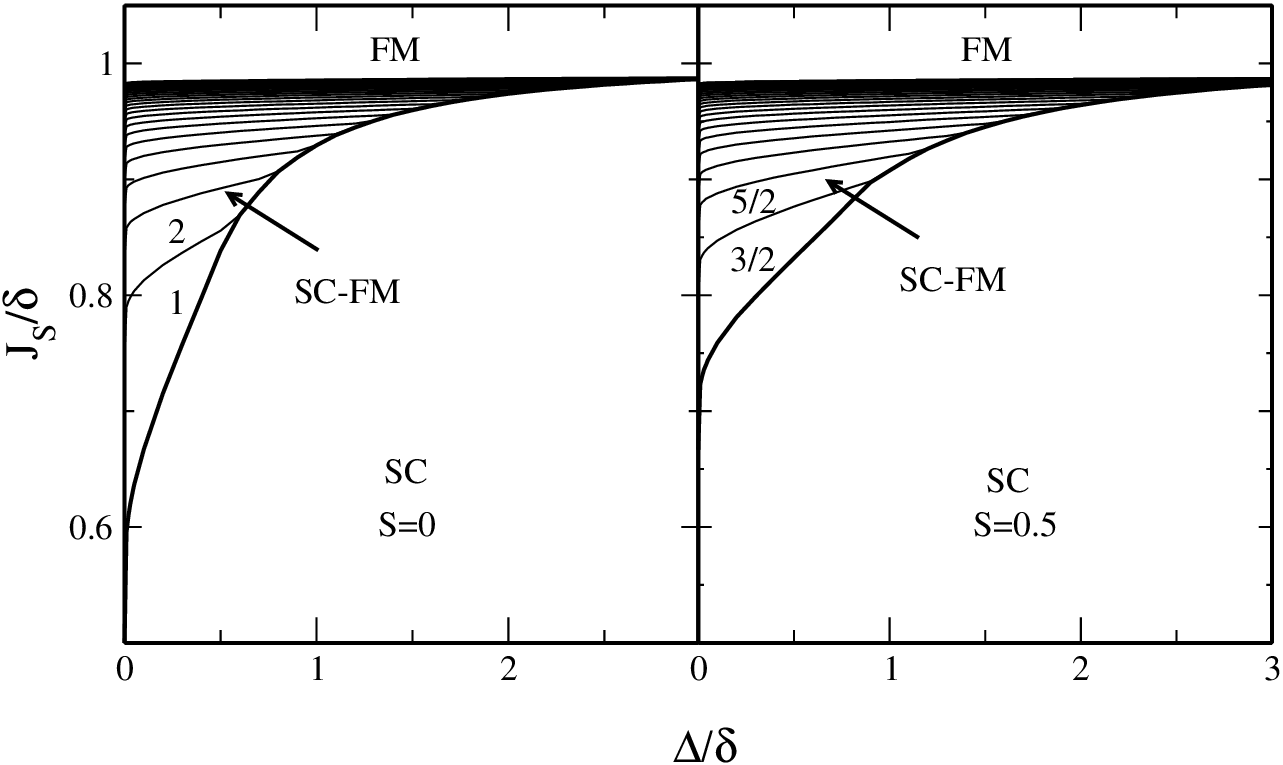}}
\caption{ Ground-state phase diagram of a grain with an equally spaced single-particle spectrum in the $J_s/\delta$--$\Delta/\delta$ plane for an even (left panel) and odd (right panel) number of electrons. The lines separate sectors of different spin values (the numbers are the corresponding spin values). We observe an intermediate regime (SC-FM) in which the ground state of the grain is partly polarized and partly paired. Reproduced from Ref.~\cite{Schmidt2007}.}
\label{fig_phase}
\end{figure}
\begin{figure}[t]
\centerline{\includegraphics[width=1.9in]{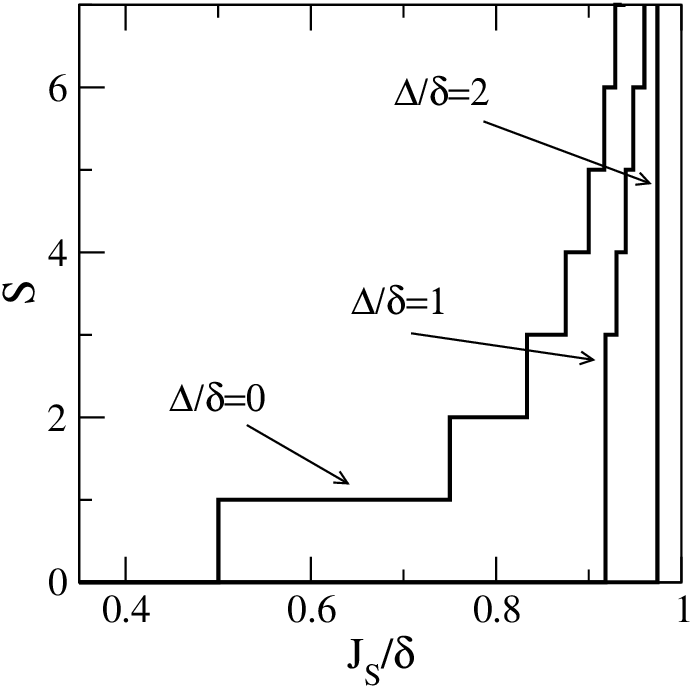}}
\caption{Ground-state spin versus the exchange coupling $J_s/\delta$ for an even grain with an equally spaced single-particle spectrum at $\Delta/\delta=0, 1, 2$. As $\Delta/\delta$ increases, the staircase function gets more compressed and pushed to higher values of $J_s/\delta$. It also exhibits spin jumps, e.g., $\Delta S=3$ for $\Delta/\delta=1$.}
\label{fig_jumps}
\end{figure}

Results for a generic equidistant single-particle spectrum with level spacing $\delta$ are shown in Fig.~\ref{fig_phase}. We find three different phases: a superconducting phase where the number of pairs is maximal, a ferromagnetic phase where the system is fully polarized with $S=N/2$, and an intermediate regime where exchange and pairing correlations coexist. The coexistence regime describes a partially polarized state, in which $2S$ electrons reside in singly occupied levels closest to the Fermi energy and the remaining electrons are paired to give spin zero \cite{Ying2006,Schmidt2007}. The existence of pairing correlations in this intermediate regime is reflected in the shift of the spin transition lines to higher values of the exchange interaction strength as the pairing gap $\Delta/\delta$ is increased.
Fig.~\ref{fig_phase} demonstrates the stepwise increase of the ground-state spin from its minimal value $S=0$ (even grain) or $1/2$ (odd grain) to its maximal value of $S=N/2$ as we increase the exchange coupling constant $J_s/\delta$ at fixed $\Delta/\delta$.

In the absence of pairing correlations (i.e., $\Delta=0$) and for an equidistant single-particle spectrum, the transition from spin $S$ to spin $S+1$ occurs for an exchange coupling of $J_s/\delta =(2S+1)/(2S + 2)$.  This stepwise increase is known as the mesoscopic Stoner staircase~\cite{Kurland2000} and is shown in Fig.~\ref{fig_jumps}.
An interesting qualitative change in the presence of pairing correlations is the possibility of a spin jump (i.e., $\Delta S >1$) for the first step. All subsequent steps have $\Delta S=1$.  For $\Delta/\delta < 0.6$, the first step corresponds to $\Delta S=1$.  However, for  $0.6 < \Delta/\delta < 0.8$, the ground-state spin changes from $S=0$ to $S=2$ in one step for a value of the exchange coupling in the range $0.87< J_s/\delta < 0.9$.  The height of the first step gets larger with increasing $\Delta/\delta$. For example, in Fig.~\ref{fig_jumps} the spin increases from $S=0$ to $S=3$ in the first step at $\Delta/\delta=1$ and then continues to increase by one unit at a time.

\subsection{Mesoscopic fluctuations}\label{spin-fluctuations}

\begin{figure}[t]
\centerline{\includegraphics[width=3in]
 {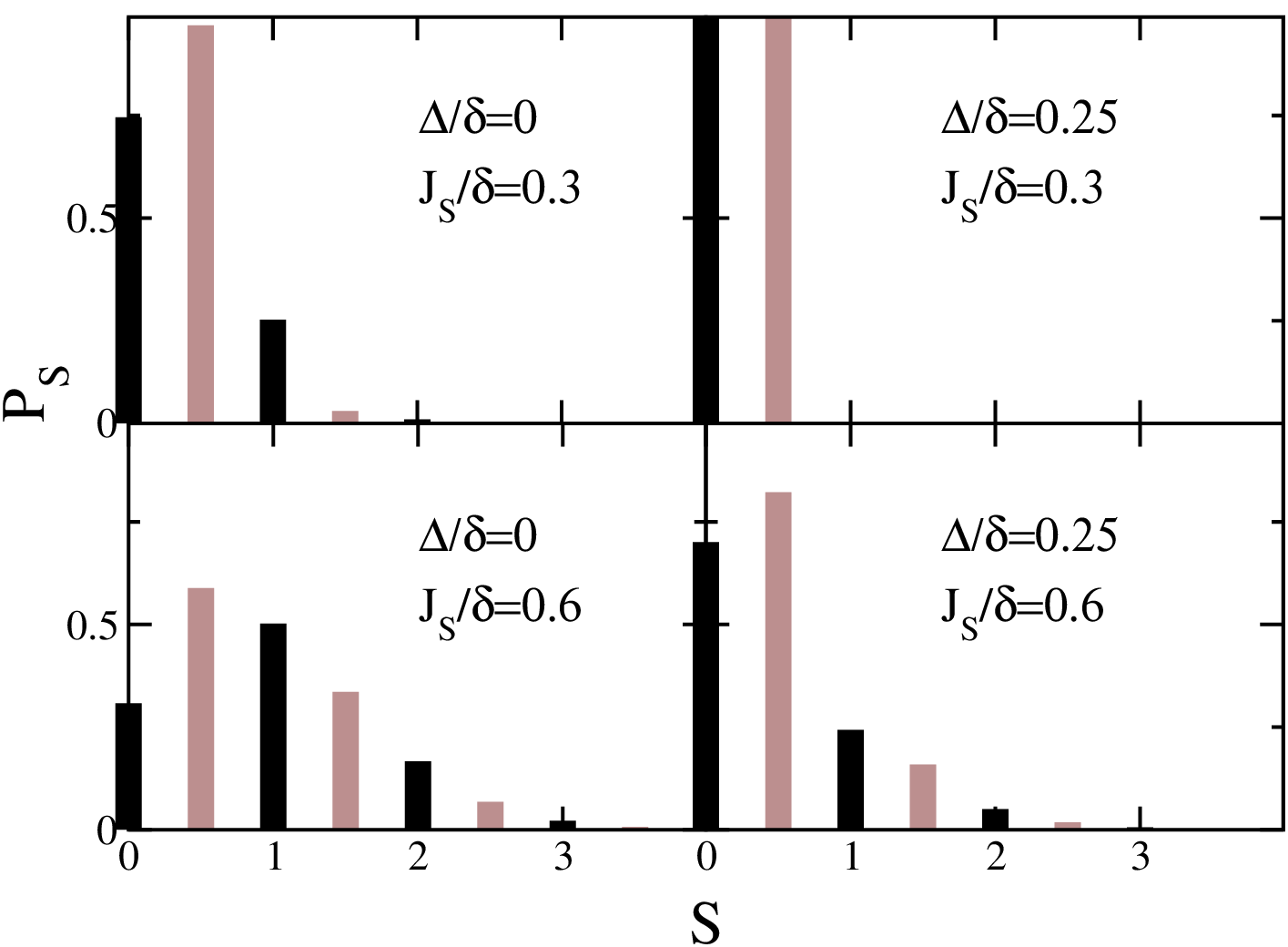}}
\caption{Ground-state spin distribution $P_S$ for a metallic grain with different values of  $\Delta/\delta$ and $J_s/\delta$. The darker (lighter) histograms describe the even (odd) grains.}
\label{fig_spindist}
\end{figure}
\begin{figure}[t]
\centerline{\includegraphics[width=2in]
 {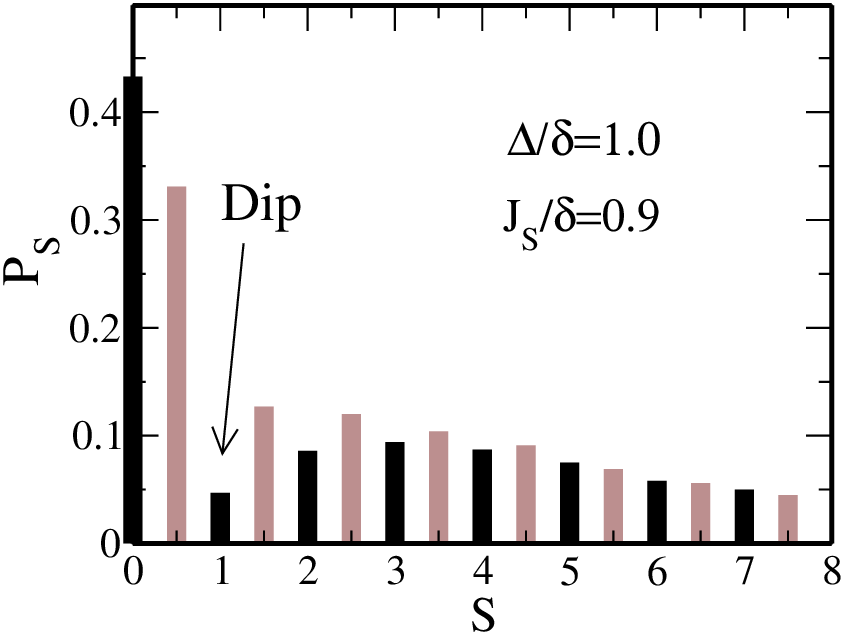}}
\caption{Ground-state spin distribution $P_S$ for a metallic grain with $\Delta/\delta=1$ and $J_s/\delta=0.9$. Results are shown for both even (darker histograms) and odd (lighter histograms) particle number. Spin jumps observed in the ground-state phase diagram of Fig.~\ref{fig_phase} (for a grain with an equidistant single-particle spectrum) manifest as a dip in the mesoscopic probability distribution.}
\label{fig_coex}
\end{figure}

The randomness of the single-particle part of the universal Hamiltonian leads to mesoscopic fluctuations of the ground-state spin. Features of the resulting ground-state spin distribution $P_S$ were studied in Ref.~\cite{Falci2003}.

Fig.~\ref{fig_spindist} shows the ground-state spin distribution for several values of the parameters $\Delta/\delta$ and $J_s/\delta$.  In the absence of interactions, the ground-state spin will always be minimal because of the cost to promote an electron out of the Fermi level to the next unoccupied level (to form a larger spin state).  This energy cost undergoes mesoscopic fluctuations (because of the GOE nature of the single-particle spectrum), and at finite exchange interaction can sometimes be compensated by the gain in exchange energy (as the spin increases).  Grains with single-particle levels close to the Fermi energy
can acquire a non-minimal ground-state spin. However, in grains with no levels sufficiently close to the Fermi energy,  the gain in exchange energy cannot compensate the cost to promote electrons to higher levels, and the ground-state spin remains minimal. As a result the spin distribution $P_S$ acquires a nonzero variance. This variance increases with the exchange coupling constant because higher ground-state spin values become more probable. The pairing interaction counteracts this behavior and favors smaller values for the ground-state spin, i.e., it suppresses
the probability for large spin values. This can be clearly seen when comparing the left ($\Delta/\delta=0$) and right ($\Delta/\delta=0.25$) columns in Fig.~\ref{fig_spindist}. Similar results for the mesoscopic ground-state spin distribution were found in Ref.~\cite{Falci2003}.

A particularly interesting signature of the mesoscopic interplay of ferromagnetism and superconductivity in chaotic metallic grains is shown in Fig.~\ref{fig_coex}. In this figure we show the spin distribution at a relatively large exchange interaction strength $J_s/\delta=0.9$ (close to the Stoner instability) and a moderate pairing strength $\Delta/\delta=1$. This case is within the coexistence regime in the ground-state phase diagram for an equidistant spectrum (see Fig.~\ref{fig_phase}) and in the vicinity of a large spin jump. We observe in Fig.~\ref{fig_coex} that the even-grain spin distribution has a local maximum at $S=3$ (besides the maximum at $S=0$) and a dip at $S=1$. Approximately $43\%$ of the grains have spin $S=0$ and $5\%$ have spin $S=1$, but $9\%$ have spin $S=3$. Thus the probability for certain spin values is anomalously suppressed. We interpret such a dip in the ground-state spin distribution as a mesocopic signature of spin jumps. Such a dip has a width of about $\Delta S\sim 1$ and occurs only at values of the exchange interaction strength that are close to the Stoner instability.

\section{Transport through a weakly coupled grain}\label{sec:Transport}
In this section we investigate the signatures of the coexistence of superconductivity and ferromagnetism in experimentally measurable quantities, i.e., the tunneling conductance of a grain that is coupled to external leads. We assume weak coupling between the grain and the leads and consider the sequential tunneling regime, where  $\delta$ and the thermal energy associated with temperature $T$ are much larger than a typical tunneling width $\Gamma$, i.e, $\delta,  T\gg\hbar\Gamma$ (here and in the following we set the Boltzmann constant $k_B=1$). In this regime the coherence between the leads and the grain can be ignored, and the conductance can be calculated using a rate equation approach~\cite{Beenakker1991,Alhassid2004}. The input in this approach are the many-particle energies and wave functions of the universal Hamiltonian, and here we obtain them using an exact diagonalization method. When the charging energy is much larger than the thermal energy, i.e., $e^2/C \gg T$, the conductance displays a series of sharp peaks as a function of gate voltage. This is the so-called Coulomb blockade regime (for reviews see, e.g., Refs.~\cite{Alhassid2000,Aleiner2002}). We studied the statistics of these conductance peaks in the quantum regime at a typical experimental temperature of $T\sim 0.1\,\delta$. For each realization of the one-body Hamiltonian, we calculated the five lowest many-particle eigenstates of the Hamiltonian (\ref{universal_hamiltonian}) using the Lanczos method for a bandwidth of $N_{\mathrm{sp}}=17$ single-particle levels and up to $N=19$ electrons. For each tunneling event, the maximal value of the linear conductance as a function of gate voltage is found numerically. We used $\sim 4000$ samples to collect sufficiently good statistics.
\subsection{Conductance peak spacings}
\begin{figure}[t]
\centerline{\includegraphics[width=2.5in]{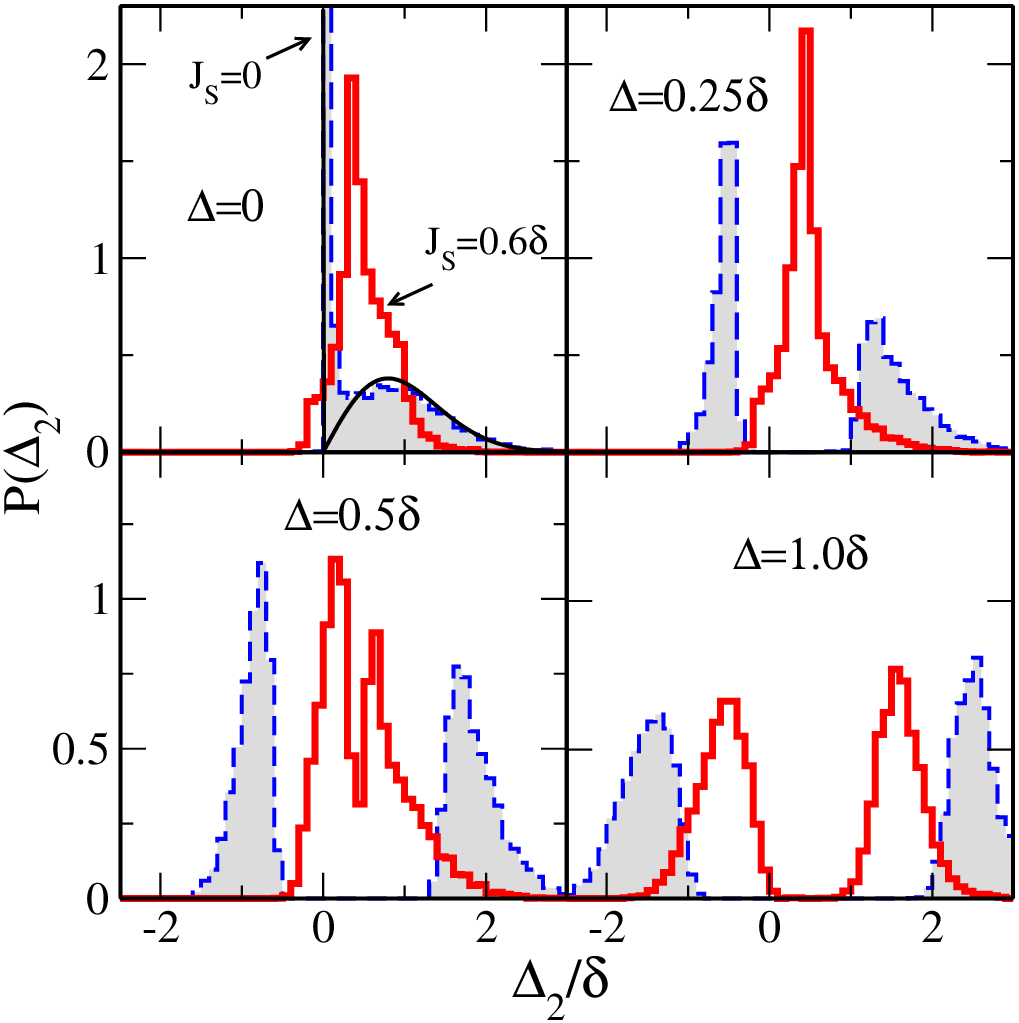}}
\caption{Peak-spacing distributions obtained from exact diagonalization at $T/\delta=0.1$ for several values of $\Delta/\delta$ shown in the panels. For each value of $\Delta/\delta$ we show results for $J_s=0$ (dashed, grey-filled histograms) and $J_s/\delta =0.6$ (solid histograms). For $\Delta=0$ we also compare with the analytic result at $T \ll \delta$ and $J_s=0$ (solid line)~\cite{Alhassid2000}. Reproduced from  Ref.~\cite{Schmidt2008}.}
\label{fig_spT010}
\end{figure}
When the total energy of a grain is approximated by the classical charging energy $e^2 N^2/2C$, the Coulomb blockade peaks are equally spaced with a spacing of $e^2/C$. The actual spacing fluctuates and is given by $e^2/C +\Delta_2$, where $\Delta_2$ incorporates the effects of the discrete single-particle spectrum and interactions beyond the classical charging energy. The quantity $\Delta_2$ is known as the peak spacing and below we discuss its statistics. 

The distributions of $\Delta_2$  are shown in Fig.~\ref{fig_spT010} for various values of $\Delta/\delta$ and $J_s/\delta$. For $\Delta=J_s=0$ and low temperatures $T \ll \delta$, the peak spacing distribution is bimodal because of a number-parity effect caused by the spin degeneracy of the single-particle levels. The exchange interaction suppresses this bimodality~\cite{Usaj2001,Alhassid2002} since it leads to fluctuating spin polarization as discussed in Sec.~\ref{spin-fluctuations}. At $J_s/\delta=0.6$, the bimodality is completely washed out.

The most significant effect of pairing correlations on the peak-spacing distribution is to restore this bimodality. For an exchange coupling of $J_s/\delta=0.6$, bimodality starts to reappear for $\Delta/\delta = 0.5$.  At even stronger pairing $\Delta/\delta=1$, the peak spacing distribution is well separated into two components.  The left part describes the sequence of even-odd-even (EOE) tunneling events, while the right part corresponds to odd-even-odd (OEO) transitions.
For $\Delta/\delta \geq 1$ the ground-state spin of an even (odd) grain is almost always $S=0$ ($S=1/2$), and mesoscopic fluctuations of the spin can be ignored. Increasing the exchange coupling constant merely shifts the two components of the distribution closer together.
The separation of the peak-spacing distribution into two parts in the presence of pairing correlations can be understood qualitatively using a fixed-$S$ BCS mean-field approximation~\cite{vonDelft2001}. In an EOE sequence, the first peak in the conductance corresponds to the blocking of an additional single-particle level and the second peak to the removal of this blocked level by creating an additional Cooper pair. In an OEO sequence, these two events are reversed, leading to a separation of the peak spacings by about $\delta\Delta_2\approx 4\Delta-3J_s$ (assuming $\Delta\gg \delta,J_s$)~\cite{Schmidt2008}. The contribution of the exchange energy is $3J_s$ since in this regime of strong pairing correlations the ground-state spin is almost always $S=0$ ($S=1/2$) for the even (odd) grain.

\subsection{Conductance peak heights}
\begin{figure}[t]
\centerline{\includegraphics[width=2.5in] {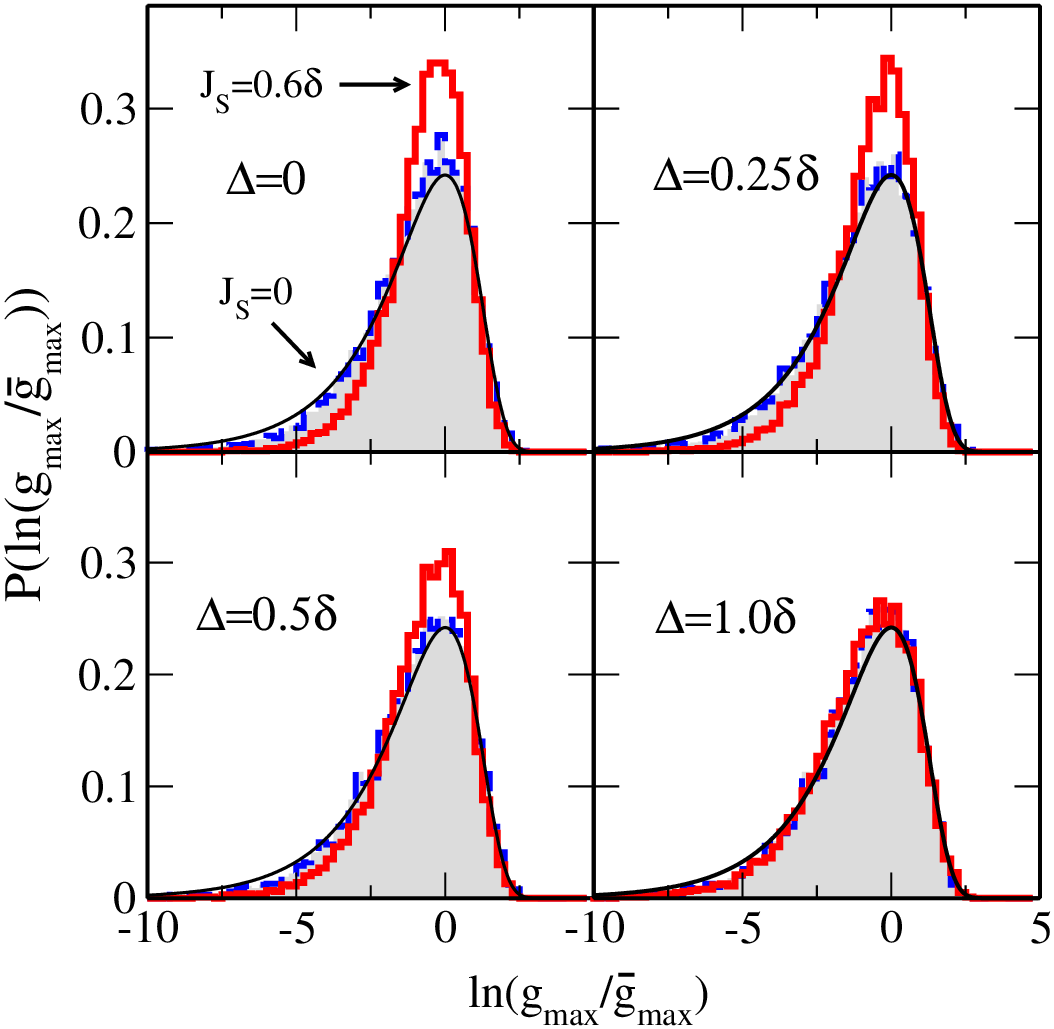}}
\caption{Peak-height distributions obtained from exact diagonalization at $T=0.1\delta$ for $J_s=0$ (dashed, grey-filled histograms) and $J_s/\delta=0.6$ (solid histograms). Different panels describe different values of $\Delta/\delta$.  We compare with the analytic result at $T\ll\delta$ and $J_s=0$ (solid line)~\cite{Jalabert1992}. Since $g_{\rm max}$ fluctuates over several orders  of magnitude, we use $\ln(g_{\rm max}/\bar g_{\rm max})$ where $\bar g_{\rm max}$ is the average conductance peak height. Reproduced from  Ref.~\cite{Schmidt2008}.}
\label{fig_phT010}
\end{figure}
\begin{figure}[t]
\centerline{\includegraphics[width=3in]{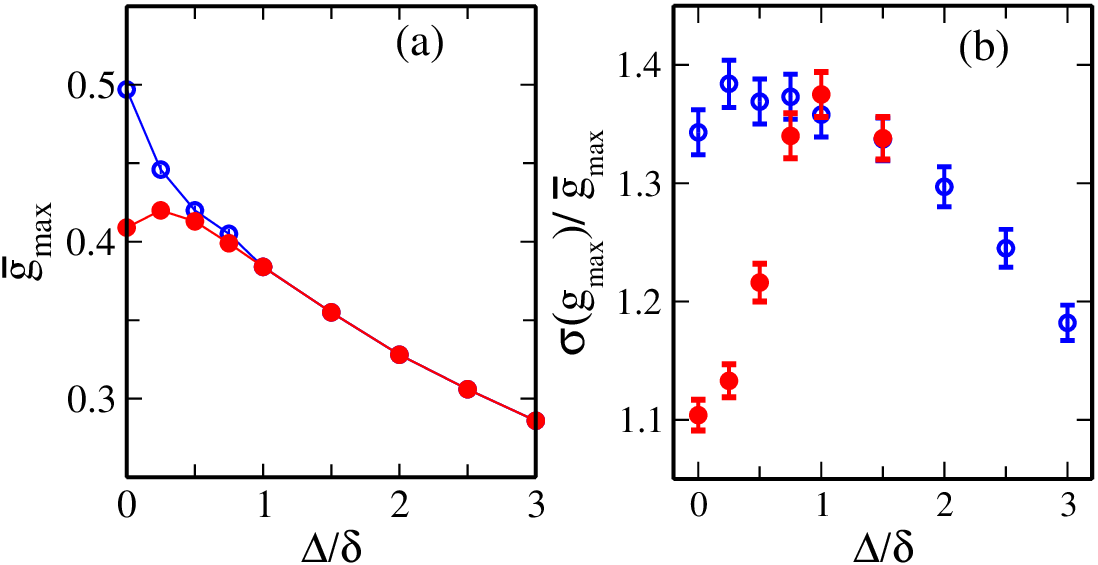}}
\caption{Average (left panel) and relative width (right panel)~\cite{Schmidt2008} of the peak height distribution as a function of $\Delta/\delta$ at $T/\delta=0.1$ for $J_s=0$ (open circles) and $J_s/\delta=0.6$ (solid circles) obtained from exact diagonalization. In the left panel we compare with the results of an equidistant single-particle spectrum (solid lines). The vertical bars in the right panel show the sampling error of the GOE distribution with 4000 samples (the errors in the left panel are smaller than the symbols). The right panel is adapted from \cite{Schmidt2008}.}
\label{fig_avstph}
\end{figure}
Distributions of conductance peak heights $g_{\rm max}$ are shown in Fig.~\ref{fig_phT010}. For $\Delta=J_s=0$ and very low temperatures $T \ll \delta$,  the peak height distribution follows from the Porter-Thomas distribution of eigenvector components of a random matrix and is known analytically ~\cite{Jalabert1992}. The exchange interaction increases the number of states contributing to the conductance by bringing down higher spin states.  This results in a narrower peak height distribution.

The mean value $\bar{g}_{\rm max}$ and the standard deviation (relative to the mean) $\sigma(g_{\rm max})/\bar{g}_{\rm max}$ of the peak height $g_{\rm max}$ are shown in Fig.~\ref{fig_avstph}. For small pairing strengths both quantities are strongly suppressed by the exchange interaction. This suppression was observed in semiconductor quantum dots, where the Cooper pairing channel can be ignored and a closed solution for the conductance is available~\cite{Alhassid2003}. There, the finite-temperature suppression of the probability of small conductance peak heights induced by exchange correlations led to a good agreement between theory and experiment at low temperatures~\cite{Alhassid2003,Usaj2003}.

The effect of pairing correlations on the excitation spectrum of the grain is to induce a gap that pushes states with large spins to higher energies. Thus, we observe that for $J_s/\delta =0.6$ the probability of small conductance peak heights and the standard deviation of the peak height distribution increase with $\Delta/\delta$ up to $\Delta/\delta \sim 1$. In particular, if the pairing interaction is sufficiently strong to destroy all spin polarization, i.e., at $\Delta/\delta=1.0$, the peak height distribution becomes independent of the exchange interaction strength $J_s/\delta$ (see Fig.~\ref{fig_phT010}).

We define a mesoscopic coexistence regime of superconductivity and ferromagnetism by the simultaneous occurrence of a bimodality in the peak spacing distribution (a signature of pairing correlations) and the suppression of peak height fluctuations (a signature of ferromagnetic correlations). In Figs.~\ref{fig_spT010}, \ref{fig_phT010} and \ref{fig_avstph} we observe such a mesoscopic coexistence case for $\Delta/\delta=0.5$ and $J_s/\delta=0.6$ at a typical experimental temperature of $T=0.1\,\delta$. Interestingly, for an equidistant spectrum this grain still belongs to the superconducting regime (see Fig.~\ref{fig_phase}). We conclude that mesoscopic fluctuations enlarge the parameter regime for which coexistence of pairing and exchange correlations can be measured. A good material for observing the coexistence regime is platinum,  which is known to be superconducting in granular form~\cite{Konig1999} and has a relatively large exchange interaction strength of $J_s/\delta \sim 0.59 - 0.72$~\cite{Gorokhov2004}.

\section{Thermodynamics}\label{sec:Thermodynamics}

\begin{figure}[t]
\includegraphics[height=190pt]{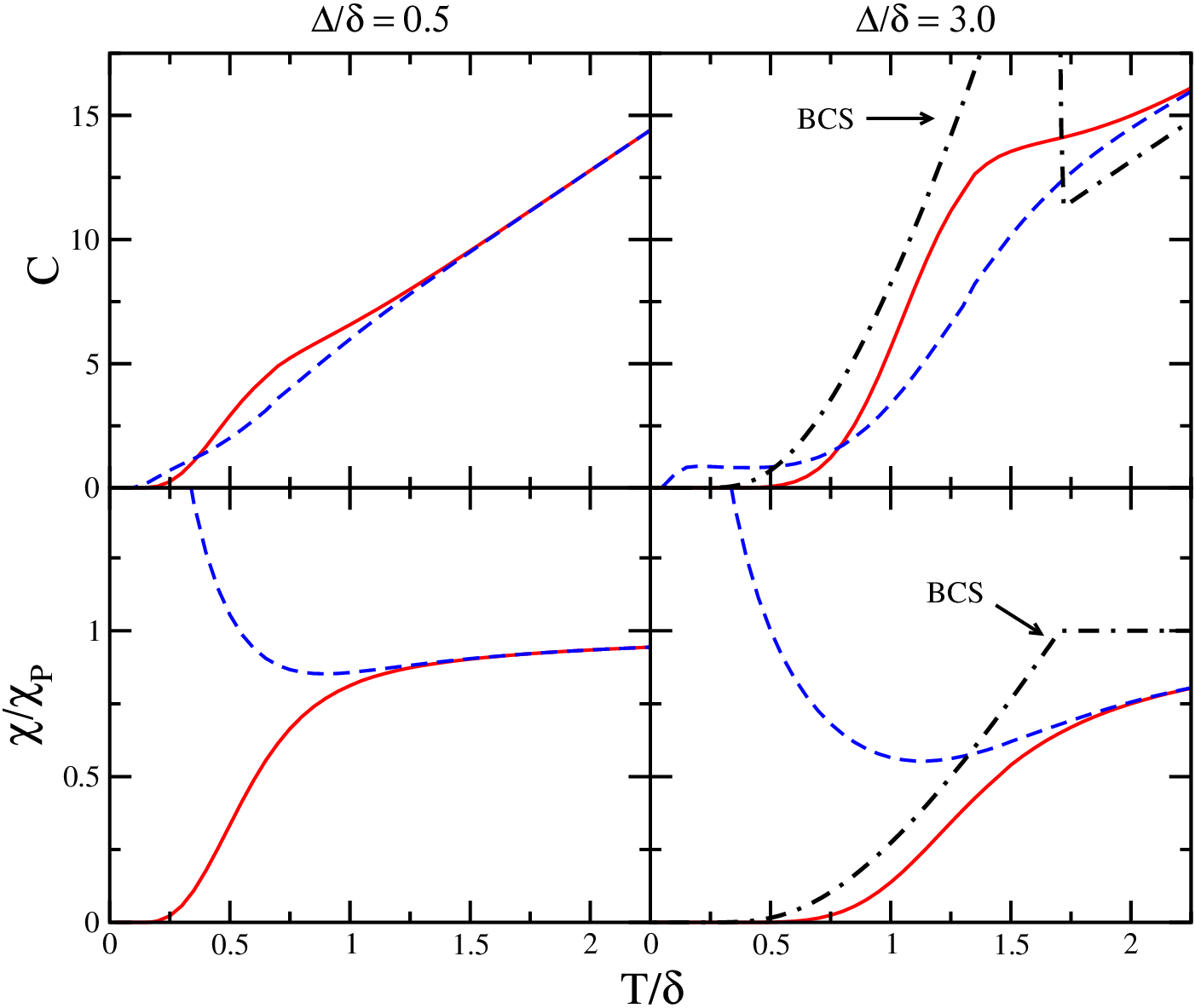}
\caption{Heat capacity (top row) and spin susceptibility (bottom row) versus temperature. The spin susceptibility is measured in units of the Pauli susceptibility $\chi_P = 2\mu_B^2/\delta$, where $\mu_B$ is the Bohr magneton. Shown are results for superconducting grains with an equally spaced single-particle spectrum, and with $\Delta/\delta=0.5$ (left column) and $\Delta/\delta=3.0$ (right column). There are no exchange correlations (i.e., $J_s=0$).  Solid and dashed lines describe grains with even and odd particle number, respectively. The observables are calculated using Richardson's solution~\cite{Richardson1963, Richardson1967} at low temperatures and the method of Ref.~\cite{Nesterov2012} at higher temperatures. The dash-dotted lines (for $\Delta/\delta=3$) are the grand-canonical BCS results. }
\label{Fig_noexchange}
\end{figure}

Here we discuss signatures of the interplay between superconductivity and ferromagnetism in thermodynamic observables of the grain.

Effects of pairing correlations on the heat capacity and spin susceptibility of ultra-small metallic grains have been studied theoretically using a number of methods~\cite{Muhlschlegel1972, DiLorenzo2000, Falci2000, Falci2002, VanHoucke2006, Alhassid2007, Schechter2001, Gladilin2004}. Thermodynamic observables of small metallic clusters were studied experimentally in Ref.~\cite{Volokitin1996}.  Signatures of pairing correlations in the bulk limit $\Delta/\delta \gg 1$ are well described by the grand-canonical BCS theory. These signatures are the exponential suppression of both the heat capacity and spin susceptibility at low temperatures because of the gap in the density of states, and a sharp peak in the heat capacity near the critical temperature $T_c$, which is a signature of a second-order phase transition (see the BCS results in Fig.~\ref{Fig_noexchange}).

In a small isolated grain, the fluctuations of the pairing gap and the dependence of observables on the parity of particle number (i.e., odd-even effects) become important~\cite{DiLorenzo2000, Falci2000,Schechter2001,Falci2002, Gladilin2004,VanHoucke2006, Alhassid2007}. An example is shown in Fig.~\ref{Fig_noexchange}, where the heat capacity (top row) and spin susceptibility (bottom row) for the pairing model with an equally spaced single-particle spectrum are shown versus temperature for $\Delta/\delta = 0.5$ (left column) and $\Delta/\delta = 3.0$ (right column). The results  for even (odd) grains are shown by solid (dashed) lines, while the grand-canonical BCS results are shown by the dash-dotted lines (for $\Delta/\delta=3.0$). Finite-size effects result in smoothing of the sharp peak in the even-grain heat capacity and its gradual transformation into a shoulder as $\Delta/\delta$ decreases. This shoulder disappears when $\Delta/\delta < 1$, but the heat capacity continues to be enhanced for an even grain (as compared with an odd grain) at intermediate temperatures. The unpaired electron in the odd grain leads to a Curie-like divergence of the spin susceptibility at low temperatures ($\chi \sim 1/T$). The combination of this divergence and the suppression of $\chi$ due to pairing correlations produces a re-entrant behavior (i.e., a local minimum) of the odd-grain spin susceptibility as a function of temperature~\cite{DiLorenzo2000}. This signature of pairing correlations exists even in the fluctuation-dominated regime $\Delta/\delta \lesssim 1$.

Exchange correlations affect the thermodynamic observables shown in Fig.~\ref{Fig_noexchange}. Signatures of the competition between exchange and pairing correlations in thermodynamic observables were studied in Refs.~\cite{VanHoucke2010} and \cite{Nesterov2012}. The case of an equally spaced single-particle spectrum was studied in \cite{VanHoucke2010} using a quantum Monte-Carlo method, while the general case of a fluctuating RMT single-particle spectrum was considered in \cite{Nesterov2012} using a semi-analytical method.
 The latter method is accurate and efficient, making it useful for  studying the mesoscopic fluctuations of thermodynamic observables.  It uses a spin projection technique~\cite{Alhassid2003,Alhassid2007a} to treat exactly the exchange interaction in (\ref{universal_hamiltonian}), and a functional integral approach to treat approximately the pairing part of (\ref{universal_hamiltonian}). The odd-even effect is reproduced by a number-parity projection. A similar approach was used in Ref.~\cite{Falci2002} to study odd-even effects in thermodynamic observables for a pure pairing Hamiltonian (i.e., in the absence of an exchange interaction).

Below we discuss this semi-analytic method and present the main results for the heat capacity and spin susceptibility of the grain~\cite{Nesterov2012}.

\subsection{Spin and number-parity projections}

The grand-canonical trace of an operator can be written as a sum of diagonal matrix elements of the operator over a complete set of quantum numbers $\lambda_1,\lambda_2,\ldots $. We define $\Tr_\lambda$ to be a trace that is restricted to the particular values of a subset $\lambda$ of this complete set (while all other quantum numbers are summed over). An important example is when $\lambda$ is the spin $S$ of the grain.  For an operator $\hat X$  we can write $\Tr \hat{X} = \sum_S\Tr_S \hat{X}$ when $\Tr_S$ denotes a trace over all states with fixed total spin $S$. The following identity allows us to express the trace of a scalar operator $\hat X$ at fixed spin $S$ in terms of the traces at fixed values of the spin component $\hat S_z$
\begin{equation}\label{spin-projection}
\Tr_S \hat{X} = (2S+1)\left(\Tr_{S_z = S} \hat{X} - \Tr_{S_z = S+1} \hat{X}\right) \;.
\end{equation}
A similar identity applies to canonical traces at fixed particle number $N$ and spin $S$.

 Using the identity (\ref{spin-projection}), we can write the canonical partition function $Z_N(J_s)$ of the universal Hamiltonian (\ref{universal_hamiltonian}) in the form
 \begin{multline}\label{Z^(J)}
 Z_N(J_s)  = \sum_S (2S+1) \\
 \times e^{\beta J_sS(S+1)}\left(Z_{N,S_z=S}-Z_{N,S_z=S+1}\right)\;,
\end{multline}
where $\beta=1/T$ and
\begin{equation}\label{Z_NS_z}
 Z_{N,S_z} = \mathrm{Tr}_{N,S_z} \left(e^{-\beta \HBCS}\right)
\end{equation}
is the $S_z$-projected canonical partition function of  the BCS-like Hamiltonian
\begin{equation}\label{BCS_Hamiltonian}
\HBCS = \sum_{i,\sigma=\uparrow,\downarrow} \epsilon_i c^\dagger_{i\sigma}
c_{i\sigma} - G \hP^\dagger\hP\,.
\end{equation}
A similar expression can be written for the spin susceptibility ($\mu_B$ is the Bohr magneton)
\begin{multline}\label{chi}
{\chi}= \frac{4\beta\mu_B^2}{3}\left\langle \bS^2\right\rangle = \frac{4\beta\mu_B^2}{3} \frac{1}{Z_N(J_s)} \sum_S S(S+1)(2S+1)
 \\ \times
  e^{\beta J_sS(S+1)}\left(Z_{N,S_z=S}-Z_{N,S_z=S+1}\right)\,.
\end{multline}

The heat capacity can be calculated from the partition function by taking numerical derivatives with respect to temperature. Thus the calculation of the heat capacity and spin susceptibility of a metallic grain described by (\ref{universal_hamiltonian}) reduces to the calculation of the $S_z$-projected partition function (\ref{Z_NS_z}).

The trace at fixed $S_z$ can be calculated using a discrete Fourier transform for the $S_z$ projection operator
\begin{equation}\label{P_S_z}
\hat{P}_{S_z} = \frac{1}{2S_{\rm max}+1}\sum_{m=-S_{\rm max}}^{S_{\rm max}} e^{i\phi_m (\hat{S}_z - S_z)}\,.
\end{equation}
Here $S_{\rm max}$ is the maximal possible value of the spin, and  $\phi_m = 2\pi m/(2S_{\rm max}+1)$ are quadrature points.

The canonical projection on a fixed particle number can be done using a similar Fourier transform, but it leads to cumbersome expressions in the framework of the path integral formalism described below. Important odd-even effects can be described by the simpler number-parity projection~\cite{Goodman1981,Rossignoli1998,Balian1999}
 \begin{equation}\label{P_eta_definition}
 \hP_\eta = \frac 12 \left(1+\eta e^{i\pi \hat{N}}\right)\,,
\end{equation}
where $\eta=1$ ($\eta=-1$) corresponds to even (odd) number of particles.
The canonical partition function for $N$ particles at fixed $\eta$ and $S_z$ is then evaluated by  the saddle-point approximation of the particle-number projection integral
\begin{equation}\label{saddle-point}
Z_{N,\eta,S_z} = \int\limits_{-i\pi/\beta}^{i\pi/\beta} \frac{\beta d\mu}{2\pi i} e^{-\beta \mu N} Z_{\eta,S_z}\;.
\end{equation}
Here
\begin{equation}\label{gc-partition}
Z_{\eta,S_z} = \Tr_{\eta,S_z} \left[e^{-\beta \left(\HBCS - \mu \hat{N}\right)}\right]
\end{equation}
is the grand-canonical partition function at fixed values of $\eta$ and $S_z$.

\subsection{Auxiliary-field path-integral formalism}

Here we present only the relevant definitions of the functional-integral formalism and summarize the results of the calculation of $Z_{N,\eta, S_z}$. For more details see  Ref.~\cite{Nesterov2012}.

We first discuss the evaluation of the grand-canonical partition function $Z_{\eta, S_z}$ in (\ref{gc-partition}).  Using the Hubbard-Stratonovich (HS) transformation~\cite{Stratonovich1957, Hubbard1959}, this partition function can be written as a functional integral over a complex auxiliary field $\Auxfield(\tau)$:
\begin{equation}\label{HS}
\\
Z_{\eta,S_z}=  \int {\mathcal{D}}[\Auxfield,\Auxfield^*] \Tr_{\eta,S_z}\left[{\cal T}e^{- \int\limits_0^\beta d\tau \left(\, |\Auxfield(\tau)|^2/G
\,+\,\hat{H}_{\Auxfield(\tau)}\right) }\right]\;.
\end{equation}
Here
\begin{multline}\label{H_eff(xi_1,xi_2)}
\hat{H}_\Auxfield
=\sum_i \left[\left(\epsilon_i - \mu-\frac
G2\right)\left(c^\dagger_{i\downarrow}c_{i\downarrow}+c^\dagger_{i\uparrow}c_{i\uparrow}\right)\right. \\ \left. - \Auxfield\, c^\dagger_{i\uparrow} c^\dagger_{i\downarrow} - \Auxfield^* \,c_{i\downarrow}c_{i\uparrow}+
\frac{G}{2}\right]
\end{multline}
describes a non-interacting Hamiltonian in an external time-dependent pairing field $\tilde\Delta(\tau)$ ($0 \leq \tau \leq \beta$) and ${\cal T}$ denotes time ordering. The saddle-point approximation to this integral (without the number-parity and $S_z$ projections) gives the BCS theory.

The auxiliary field $\Auxfield(\tau)$ can be separated into its static and time-dependent parts by  expanding in a Fourier series
\begin{equation}
	\Auxfield(\tau) = \Delta_0 + \sum_{r\ne 0} \Delta_r e^{i\omega_r\tau}\,,
\end{equation}
where  $\omega_r = {2\pi r}/{\beta}$ ($r$ integer) are the bosonic Matsubara frequencies. In the static-path approximation (SPA) \cite{Muhlschlegel1972, Alhassid1984, Lauritzen1988}, we ignore the time dependance of $\Delta$ (i.e., we take $\Delta_r=0$ for all $r\neq 0$) and carry out the integral over the static field $\Delta_0$ exactly. In this approximation
\begin{equation}\label{Z-SPA}
Z_{\eta, S_z} \approx  \int \limits_0^\infty \frac{\beta \,d\,\Dstatic^2}{G}e^{-(\beta/G)\Dstatic^2}   Z_{\eta, S_z}(\Dstaticarg)\,,
\end{equation}
where $Z_{\eta, S_z}(\Dstaticarg) = \Tr_{\eta,S_z} e^{-\beta \hat{H}_{\Dstaticarg}}$ is a
static partition function.  Using the explicit expressions (\ref{P_S_z}) and (\ref{P_eta_definition}) for the projection operators, and working in a $\Dstaticarg$-dependent quasiparticle basis that diagonalizes $\hat{H}_{\Dstaticarg}$, we find~\cite{Nesterov2012}
\begin{multline}\label{Z(Dstatic)}
Z_{\eta, S_z}(\Dstaticarg)= \left(\prod_i e^{-\beta(\epsilon_i - \mu-E_i)} \right)\sum_m \frac{e^{-i\phi_m S_z}}{2(2S_{\mathrm{max}}+1)}
\\
\times \left(\prod_i \left|1+e^{-\beta E_i - i\frac{\phi_m}{2}}\right|^2 + \eta \prod_i \left|1-e^{-\beta E_i - i\frac{\phi_m}{2}}\right|^2 \right)\,,
\end{multline}
where $E_i =\sqrt{\left(\epsilon_i - \mu - \frac G2\right)^2 + \Dstatic^2}$ are the quasiparticle energies.

The SPA can be improved by carrying out the integration over $\Delta_r$ (with $r \ne 0$) in the saddle-point approximation for each value of $\Delta_0$. This leads to the random-phase approximation (RPA) correction to the SPA~\cite{Kerman1981, Kerman1983, Puddu1991, Lauritzen1993, Rossignoli1997, Attias1997,Rossignoli1998}. This correction amounts to multiplying $Z_{\eta,S_z}(\Dstaticarg)$ in (\ref{Z-SPA}) by the factor
\begin{equation}\label{C_RPA = sinh/sinh}
C^{\mathrm{RPA}}_{\eta,S_z} (\Dstaticarg) = \prod_i
\frac{\Omega_i}{2E_i}\frac{\sinh(\beta
E_i)}{\sinh\left(\frac{\beta \Omega_i}2\right)}\,.
\end{equation}
Here $\pm \Omega_i$ are the eigenvalues of
the $2N_{\text{sp}}\times 2N_{\text{sp}}$ RPA matrix ($N_{\text{sp}}$ is the number of single-particle orbitals)
\begin{equation}\label{RPA_matrix}
\left(%
\begin{array}{cc}
  2E_i \delta_{ij} - \frac G2f_{i}(\gamma_i\gamma_j+1) & -\frac G2 f_{i}(\gamma_i\gamma_j-1)  \\
  \frac G2 f_{i}(\gamma_i\gamma_j-1)  & \frac G2 f_{i}(\gamma_i\gamma_j+1)- 2E_i \delta_{ij} \\
\end{array}%
\right)\,,
\end{equation}
where
\begin{equation}
\gamma_i = \frac{\epsilon_i - \mu - \frac G2}{E_i}
\end{equation}
and
\begin{equation}\label{F_eta_i}
f_{i} = \frac{1}{\beta} \frac{\partial \ln Z_{\eta,S_z}(\Dstaticarg)}{\partial E_i}\,.
\end{equation}

Next we discuss the approximate evaluation of the canonical partition function $Z_{N,\eta, S_z}$ in (\ref{saddle-point}). Substituting the HS transformation (\ref{HS}) for $Z_{\eta, S_z}$ in the integrand of  (\ref{saddle-point}) and exchanging the order of integrations, we evaluate the integral over $\mu$ in a saddle-point approximation for each fixed value of $\Dstaticarg$. This saddle-point integration is applied to the grand-canonical static free energy
\begin{equation}
F(\Dstaticarg) = \Dstatic^2/G -T\ln Z(\Dstaticarg) \;,
\end{equation}
where
\begin{equation}
Z(\Dstaticarg) = \Tr\, e^{-\beta \Hstatic} = \prod_i 4e^{-\beta(\epsilon_i-\mu)} \cosh^2 \frac{\beta E_i}{2}\;.
\end{equation}
The remaining ratio $Z_{\eta, S_z}(\Dstaticarg) C^{\mathrm{RPA}}_{\eta,S_z}(\Dstaticarg)/Z(\Dstaticarg)$ is treated as a pre-factor that does not enter in the saddle-point integration. We find
\begin{multline}\label{Z_N}
Z_{N,\eta, S_z}  \approx\int \limits_0^\infty \frac{\beta \,d\,\Dstatic^2}{G} \left(\frac{2\pi}{\beta}\left|\frac{\partial^2 F}{\partial \mu^2}\right|\right)^{-1/2} \\
\times e^{-(\beta/G)\Dstatic^2} \,\, e^{-\beta \mu N}  Z_{\eta, S_z}(\Dstaticarg) \,\,C^{\mathrm{RPA}}_{\eta, S_z}(\Dstaticarg)
\end{multline}
with
\begin{equation}\label{d2Fdmu2}
\frac{\partial^2 F}{\partial \mu^2}= - \sum_i\frac{\beta
E_i\left(\epsilon_i - \mu -\frac G2\right)^2 + \Dstatic^2\sinh(\beta E_i)
}{2E_i^3\cosh^2\left(\frac{\beta E_i}{2}\right)}\;.
\end{equation}
The saddle-point equation for $\mu$ has the usual form of the particle-number equation
\begin{equation}\label{N=-dF/dmu}
N = -\frac{\partial F}{\partial \mu}= \sum_i\left(1 -
\frac{\epsilon_i - \mu - \frac G2}{E_i}\tanh\frac{\beta
E_i}{2}\right)
\end{equation}
but its solution $\mu=\mu(\Dstaticarg)$ depends on the value of the static field.

\begin{figure}[t]
\includegraphics[height=250pt]{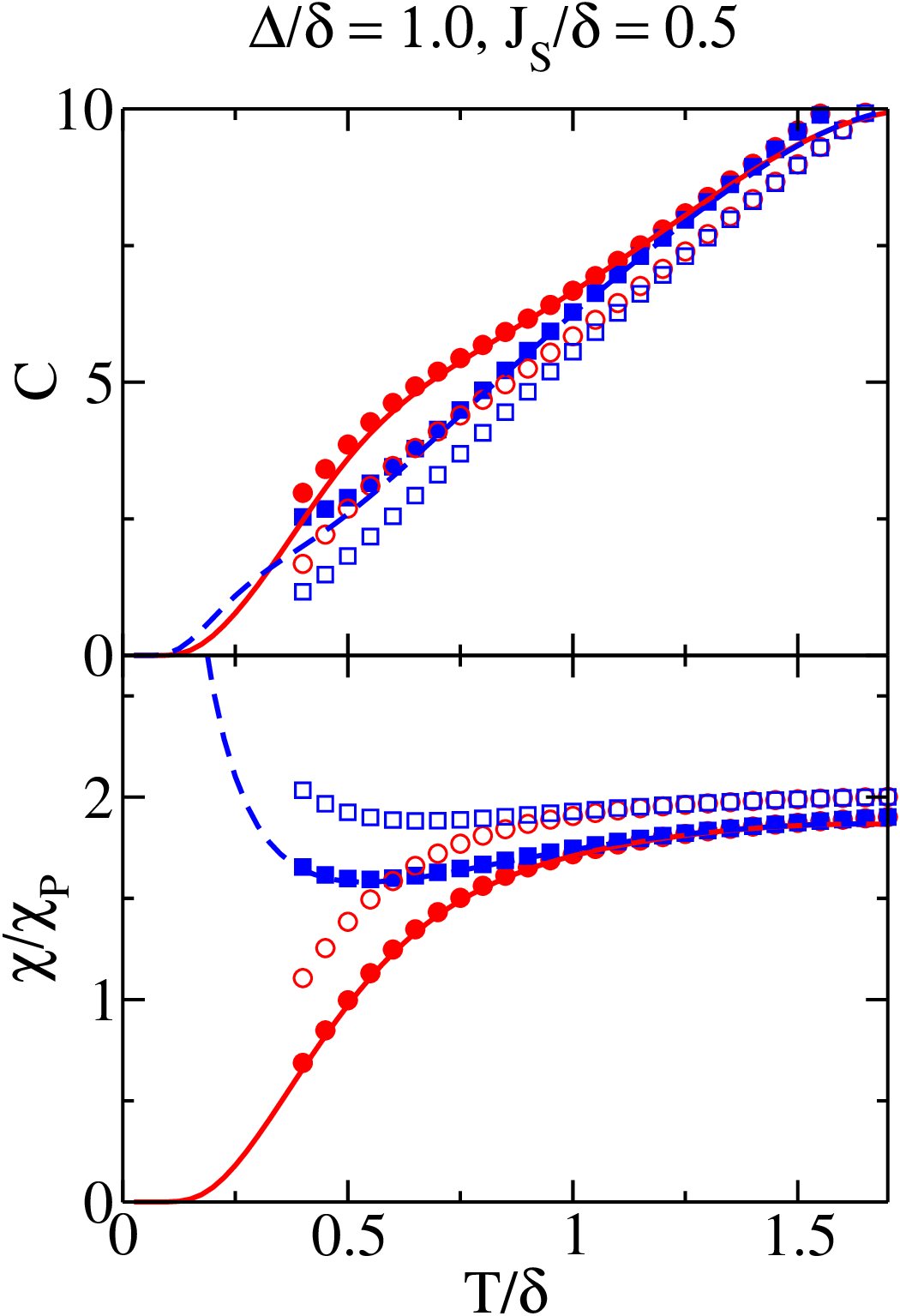}
\caption{Heat capacity $C$ (top panel) and spin susceptibility $\chi$ in units of the Pauli susceptibility $\chi_P$ (bottom panel) versus temperature $T/\delta$ of a metallic grain described by the universal Hamiltonian (\ref{universal_hamiltonian}) with equally spaced single-particle levels, and with $\Delta/\delta = 1.0$, $J_s/\delta=0.5$. Results are shown for both even particle number (solid line, circles) and odd particle number (dashed line, squares). Results obtained using the generalized  Richardson's solution of Sec.~\ref{solution} (lines) are compared with the SPA results (open symbols) and the SPA+RPA results (solid symbols).}
\label{rich}
\end{figure}

We have shown that this number-parity projected SPA+RPA method is quite accurate by comparing it with exact results. The latter were calculated from the many-body eigenvalues of (\ref{universal_hamiltonian}) using Richardson's solution generalized to include the exchange interaction (see Sec.~\ref{solution}). In Fig.~\ref{rich} we show the heat capacity (top panel) and spin susceptibility (bottom panel) for a grain with equally spaced single-particle spectrum and with $\Delta/\delta = 1$, $J_s/\delta=0.5$.  The exact results (lines) are compared with the SPA results (open symbols) and the SPA+RPA results (solid symbols) for both even and odd grains.  These results demonstrate (i) the importance of the RPA correction, and (ii) the good accuracy of the SPA+RPA calculations. We note that Richardson's solution is not practical at high temperatures since the number of required eigenvalues increases exponentially. The deviation of the exact results from the SPA+RPA results at the higher temperatures in Fig.~\ref{rich} is because the exact many-body eigenvalues were calculated only below a cutoff of $\sim 30\,\delta$.

The SPA+RPA method is not applicable below a certain critical temperature, when the gaussian integration over one of the fluctuations $\Delta_r$ diverges for at least one value of $\Dstaticarg$, making the RPA correction unstable. It was recently proposed~\cite{Ribeiro2012} that this problem can be overcome by treating non-perturbatively a low-energy collective mode. At low temperatures, one can also use the exact solution (the number of eigenvalues required to calculate thermal observables at sufficiently low temperatures is not large).

\subsection{Heat capacity}

\begin{figure}[t]
\includegraphics[height=240pt]{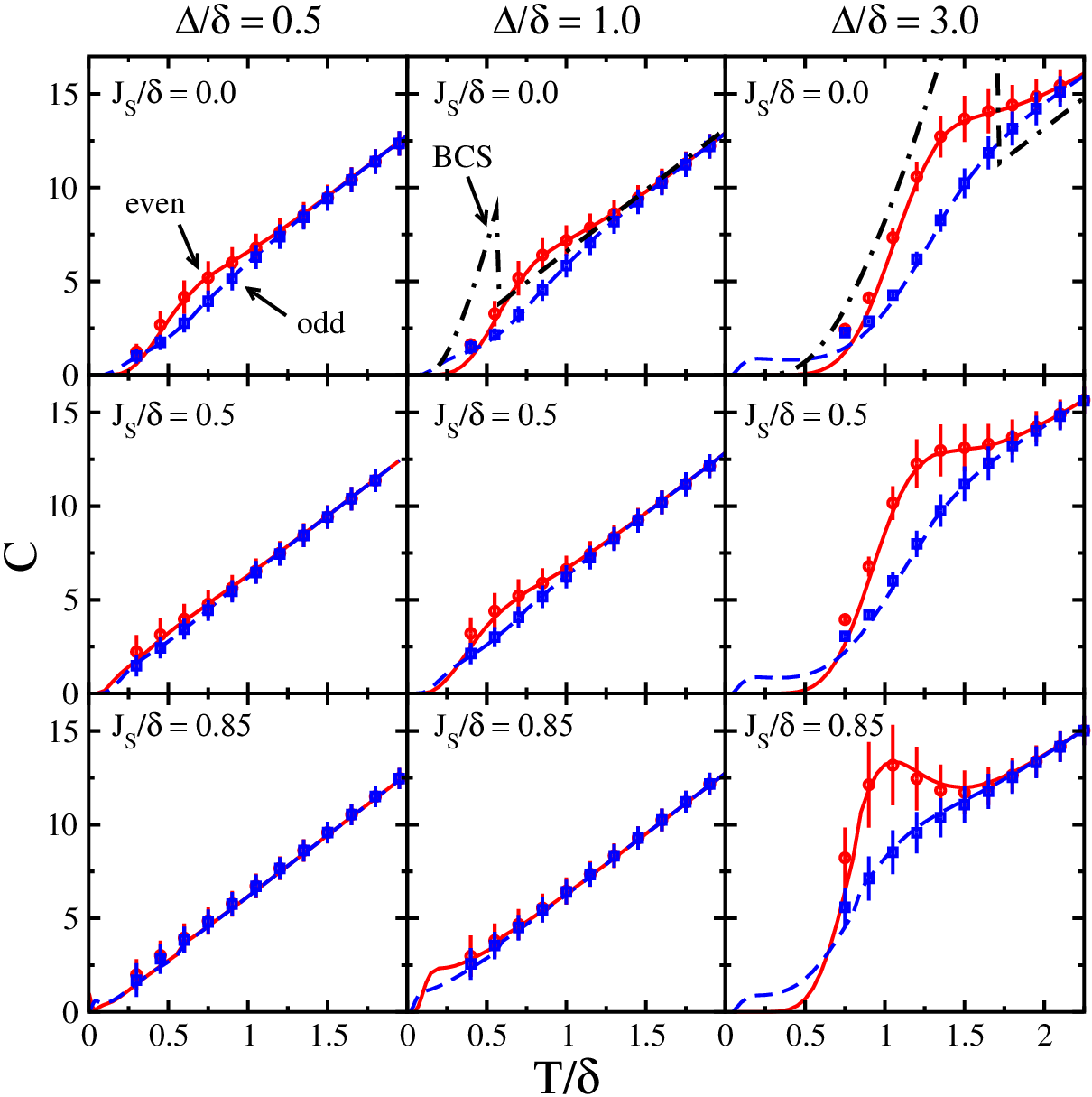}
\caption{The heat capacity $C$  versus temperature $T/\delta$ for an even grain (solid lines, circles) and for an odd grain (dashed lines, squares). Results are shown for grains with
$\Delta/\delta=0.5$ (left column), $\Delta/\delta=1.0$ (middle column) and
$\Delta/\delta=3.0$ (right column), and with $J_s/\delta=0$ (top row), $J_s/\delta=0.5$ (middle row) and $J_s/\delta=0.85$ (bottom row). The
symbols and vertical bars describe, respectively, the average value $\overline{C}$ and standard deviation $\delta C$ of the heat capacity for a fluctuating RMT single-particle spectrum in (\ref{universal_hamiltonian}).  The lines correspond to an equally spaced single-particle spectrum, and the dash-dotted lines are the grand-canonical BCS results (where applicable).}\label{hc_mes}
\end{figure}

\begin{figure}[t]
\includegraphics[height=240pt]{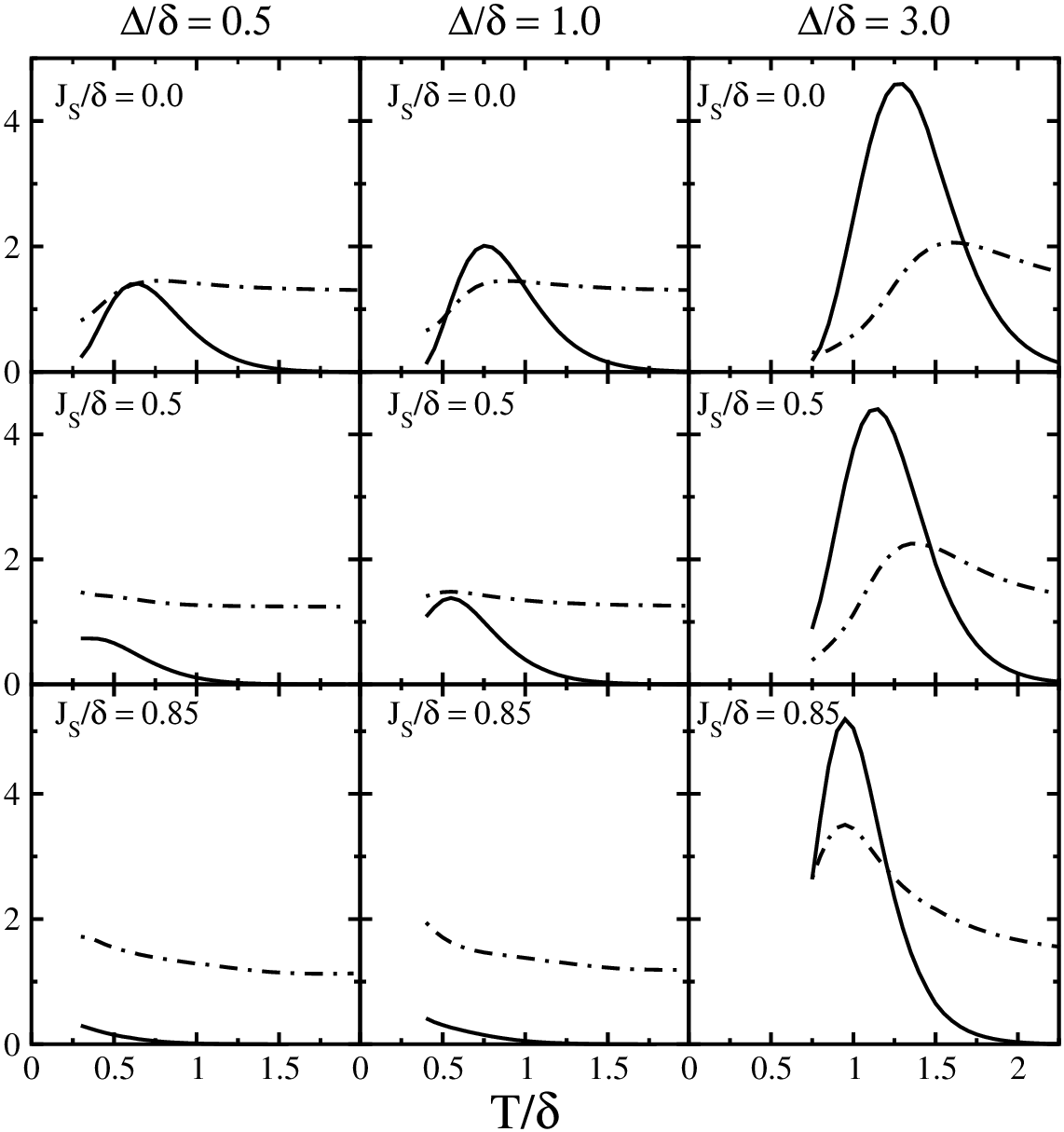}
\caption{The sum $\delta C_{\text{even}} + \delta C_{\text{odd}}$ of standard deviations (dash-dotted lines) and the difference $\overline{C}_{\text{even}} - \overline{C}_{\text{odd}}$ between average values (solid lines) of the heat capacities of an even grain and an odd grain  are shown versus temperature.}
\label{hc_evenodd}
\end{figure}

The main results for the heat capacity $C$ as a function of temperature are shown in Fig.~\ref{hc_mes} for several values of $\Delta/\delta$  and $J_s/\delta$. The symbols with vertical bars are, respectively, the average values $\overline{C}$ and standard deviations $\delta C$ calculated for an ensemble of 1000 realizations of single-particle GOE spectra in (\ref{universal_hamiltonian}). The lines are the results for an equally spaced single-particle spectrum (obtained from the exact solution at low temperatures and from the SPA+RPA at higher temperatures).  The number of single-particle levels $N_{\text{sp}}$ in the model space was odd and varied between 31 and 61 (depending on temperature) and the particle number was either $N_{\text{sp}}-1$ or $N_{\text{sp}}$ (depending on its parity).
The interplay between odd-even effects and mesoscopic fluctuations is further demonstrated in Fig.~\ref{hc_evenodd}. The sum $\delta C_{\text{even}} + \delta C_{\text{odd}}$  of the standard deviations for the even and odd grains is shown by dash-dotted lines, while the difference $\overline{C}_{\text{even}} - \overline{C}_{\text{odd}}$ between the average values for the even and odd grains is shown by solid lines.

Both figures \ref{hc_mes} and \ref{hc_evenodd} show that the exchange interaction shifts the odd-even effect to lower temperatures. In addition, when $\Delta \lesssim \delta$, the exchange suppresses the odd-even effect. However, when $\Delta > \delta$, the exchange suppresses only the right side of the even-grain shoulder and therefore transforms this shoulder into a peak. Overall, the effect of exchange is less prominent when $\Delta > \delta$; higher values of $J_s/\delta$ are required to produce a visible change, and those changes occur mostly at higher temperatures.

These results are consistent with the phase diagram discussed in Sec.~\ref{sec:Phase_Diagram}. The dependence of a thermodynamic observable on $J_s/\delta$ arises through the dependence of the many-particle energies on $J_s/\delta$, which is determined by the total spin $S$ of the states. A typical excitation energy of a state with non-ground-state spin increases with $\Delta/\delta$  and thus the change in the relative weight of such a state decreases with increasing $\Delta/\delta$. At larger values of $\Delta/\delta$, this relative change becomes appreciable only at higher temperatures, resulting in the suppression of the right side of the even-grain shoulder but in a much weaker dependence on $J_s/\delta$ of its left side.

Another interesting question is whether the odd-even effect in the heat capacity can be resolved despite the mesoscopic fluctuations in experiments in which the number parity of electrons in the grain is unknown. The condition for resolving the odd-even effect in such experiments is that $\overline{C}_{\text{even}} - \overline{C}_{\text{odd}}$ be larger than $\delta C_{\text{even}} + \delta C_{\text{odd}}$, namely, when the solid line in Fig.~\ref{hc_evenodd} is above the dash-dotted line. We observe that in larger grains with $\Delta/\delta > 1$, there is always a substantial temperature interval in which this condition is satisfied. In the crossover regime $\Delta/\delta \sim 1$, the odd-even effect starts to be washed out by mesoscopic fluctuations for a moderate strength of the exchange interaction. In the smaller grains with $\Delta/\delta = 0.5$, the observability of the odd-even effect is marginal already in the absence of exchange interaction.

\subsection{Spin susceptibility}

\begin{figure}[t]
\includegraphics[height=240pt]{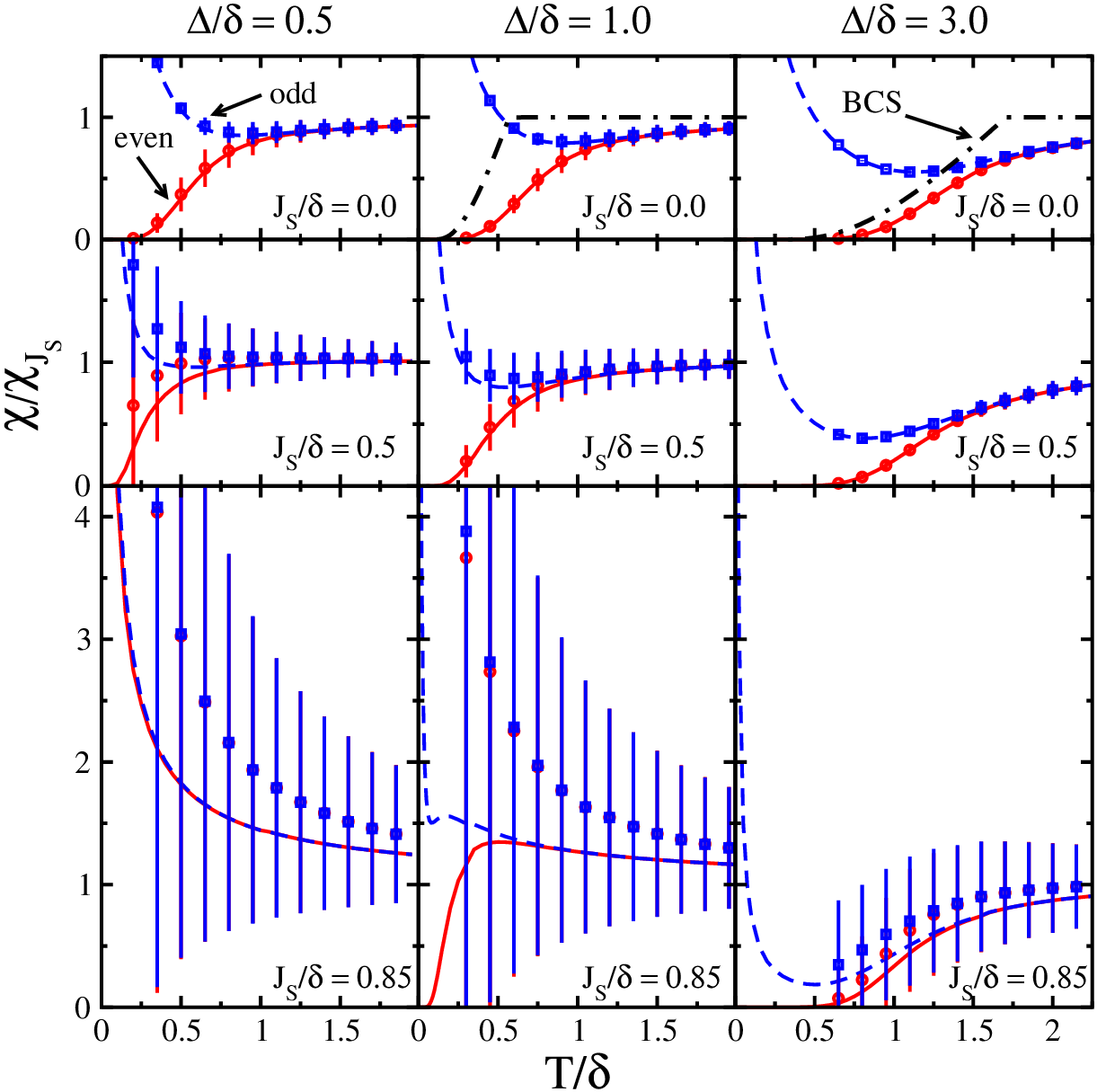}
\caption{The spin susceptibility $\chi$ in units of  $\chi_{J_s} = \chi_P/(1-J_s/\delta)$ versus temperature $T/\delta$ for even and odd grains for the same values of
$\Delta/\delta$ and $J_s/\delta$ as in Fig.~\ref{hc_mes}.  Symbols and lines follow the same convention as in Fig.~\ref{hc_mes} but for the spin susceptibility.}
\label{ss_mes}
\end{figure}

\begin{figure}[t]
\includegraphics[height=240pt]{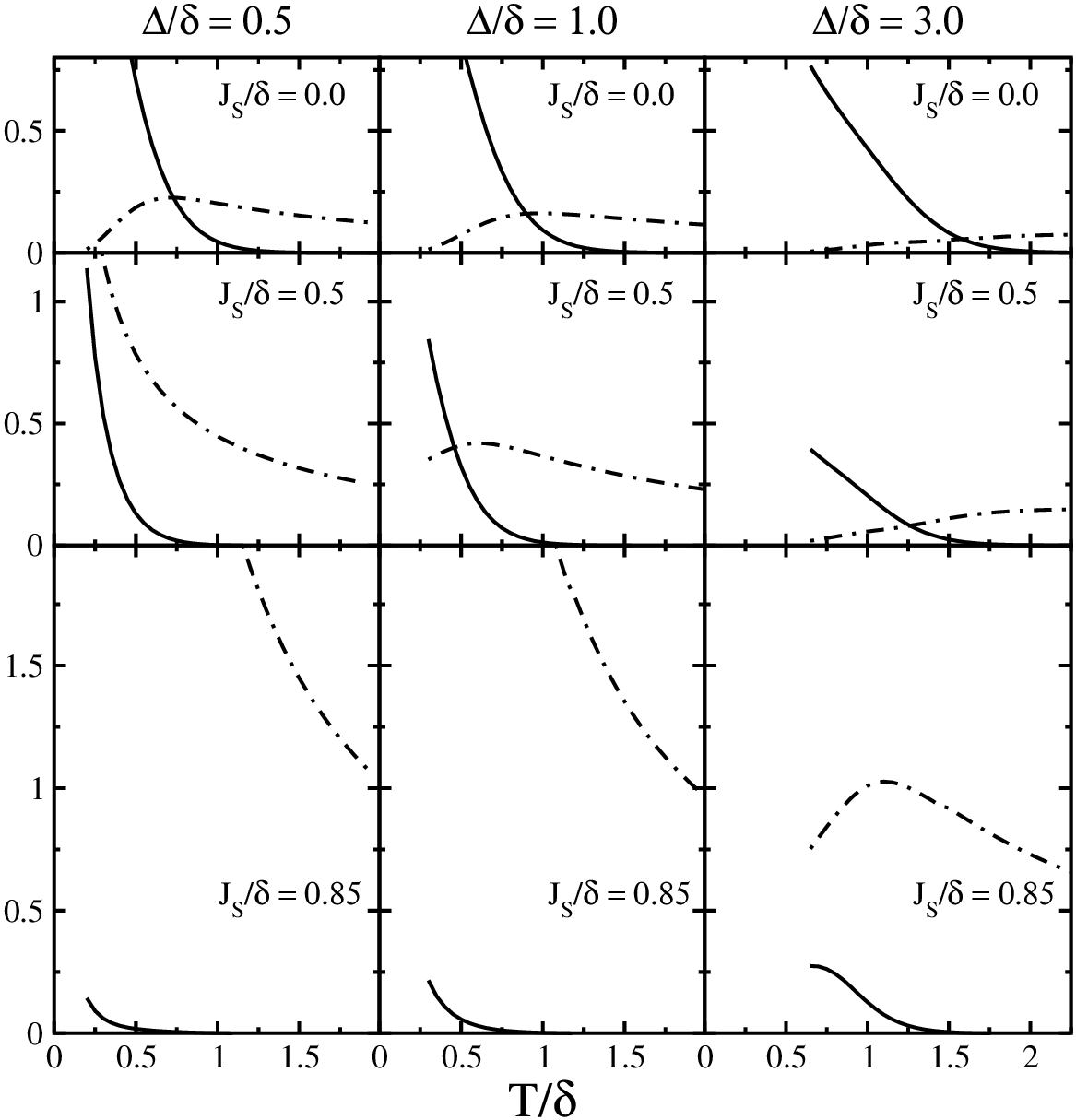}
\caption{The sum $\delta \chi_{\text{even}} + \delta \chi_{\text{odd}}$ of standard deviations  (dash-dotted lines) and the difference $\overline{\chi}_{\text{odd}} - \overline{\chi}_{\text{even}}$ between average values (solid lines)
of the spin susceptibilities for an odd grain and for an even grain versus temperature. The susceptibility $\chi$ is measured in units of  $\chi_{J_s} = \chi_P/(1-J_s/\delta)$.}
\label{ss_evenodd}
\end{figure}

The results for the spin susceptibility $\chi$ are shown in Figs.~\ref{ss_mes} and \ref{ss_evenodd} for the same ensemble of Hamiltonians used for the heat capacity. In both figures, $\chi$ is normalized by the $J_s$-dependent high-temperature limit of its average value $\chi_{J_s} = \chi_P/(1-J_s/\delta)$~\cite{Burmistrov2010,Burmistrov2012}, where $\chi_P = 2\mu_B^2/\delta$ is the Pauli susceptibility.  Our notation follows the same convention as in Figs.~\ref{hc_mes} and \ref{hc_evenodd}, except that the solid lines in Fig.~\ref{ss_evenodd} correspond to the odd-even difference of spin susceptibilities (rather than the even-odd difference taken for the heat capacity).

As with the heat capacity, the exchange interaction can shift odd-even effect in the spin susceptibility to lower temperatures. When $\Delta/\delta \lesssim 1$, the exchange can completely suppress the re-entrant behavior of the odd-grain average spin susceptibility. However, when $\Delta/\delta > 1$, the re-entrant behavior is enhanced by the exchange interaction. This effect is similar to the suppression of the right side of the even-grain shoulder in the heat capacity, but it is more prominent for the spin susceptibility, whose high-temperature value is strongly affected by exchange correlations.

In general, exchange correlations enhance the mesoscopic fluctuations of $\chi$. These fluctuations become particularly strong (as $J_s/\delta$ increases) in the fluctuations-dominated regime, even when $\chi$ is measured in the units of the already enhanced average value $\chi_{J_s}$. These strong fluctuations of $\chi$ in the smaller grains are explained by the large dispersion of the magnetization of the grain or alternatively by the strong dependence of $\chi$ on excitation energies of states with different spin values. Fluctuations in such excitation energies result in very large fluctuations of $\chi$ in the smallest grains.

\section{Conclusions}
\label{sec:Conclusions}

We have discussed various signatures of the interplay between superconducting and ferromagnetic correlations in  nano-scale metallic grains. In particular, we have considered chaotic grains with a large dimensionless Thouless conductance, which are well described by the universal Hamiltonian. The non-interacting part of this Hamiltonian is fluctuating and is modeled by random-matrix theory, while the coupling constants of its interaction part do not fluctuate. For a fixed particle number, this universal interaction consists of pairing and exchange terms describing interactions in the Cooper and spin channels, respectively. The single-particle part of the Hamiltonian and the pairing term favor minimization of the total spin and lead to superconductivity in the bulk, while the spin exchange term tends to polarize the grain and leads to ferromagnetism when the exchange coupling constant exceeds the Stoner instability threshold. We have identified signatures of the competition between pairing and exchange correlations in the following observables: (i) the ground-state spin of the grain; (ii) the mesoscopic fluctuations of the conductance of a grain that is weakly coupled to leads, and (iii) thermodynamic observables such as the heat capacity and spin susceptibility. We summarize these signatures as follows.

(i){\em  Ground-state spin}. The ground-state phase diagram of the grain exhibits a coexistence regime in which the electrons in the grain are partly paired and partly polarized. The ground-state spin at fixed $\Delta/\delta$ is a monotonic stepwise function of $J_s/\delta$. The values of $J_s/\delta$ at which such steps occur increase with $\Delta/\delta$. When $\Delta/\delta$ is sufficiently large, the  change $\Delta S$ of the ground-state spin in the first step can be larger than 1 (this is known as a spin jump).

The mesoscopic fluctuations of the ground-state spin acquire a finite variance in the presence of exchange correlations, and they are suppressed by the competing pairing correlations. When both the exchange and pairing interactions are sufficiently strong  ($\Delta/\delta \sim 1$ and $J_s/\delta \sim 0.9$), the ground-state spin distribution has a local minimum, which is a signature of spin jumps.

(ii){\em Linear conductance}. In the absence of pairing and exchange correlations, the distribution of the Coulomb-blockade conductance peak spacings is expected to be bimodal (because the single-particle energy levels are spin degenerate). In semiconductor quantum dots (in which the pairing interaction is repulsive and its effects can be ignored), this bimodality is suppressed by the exchange interaction. However, the competing attractive pairing correlations in metallic grains can restore the bimodality of the conductance peak spacing distribution.

The number of many-particle energy levels that participate in the low-temperature conductance increases in the presence of exchange correlations since levels with higher spin values shift down in energy. This leads to the suppression of the peak height fluctuations. In contrast, pairing correlations reduce the number of low-lying levels (because of the pairing gap) and thus tend to enhance the peak height fluctuations.

There is a regime in the parameter space $\Delta/\delta-J_s/\delta$ in which the peak spacing distribution exhibits the signature of pairing correlations (i.e., it is bimodal), while the peak height fluctuations show the signature of exchange  correlations (i.e., they are suppressed). This is a mesoscopic coexistence regime of pairing and exchange correlations.

(iii) {\em Thermodynamics}. Pairing correlations lead to a number-parity effect in the heat capacity and in the spin susceptibility of the grain. This effect is shifted to lower temperatures in the presence of exchange correlations. In the fluctuation-dominated regime $\Delta/\delta \lesssim 1$, the odd-even effect is suppressed by exchange correlations. However, for $\Delta/\delta >1$, certain signatures of pairing correlations, such as a peak in the heat capacity for an even grain and a local minimum in the spin susceptibility of an odd grain, are enhanced by exchange correlations. This is explained by the stronger dependence of thermodynamic observables on the exchange coupling strength at higher temperatures and the much weaker dependence at low temperatures.  Mesoscopic fluctuations of thermodynamic observables can wash out the odd-even effects for sufficiently small $\Delta/\delta$ or large $J_s/\delta$. The fluctuations of the spin susceptibility are more sensitive to exchange correlations in the fluctuations-dominated regime and become particularly large as $J_s/\delta$ approaches the Stoner instability.

\begin{acknowledgments}
This work was supported in part by the U.S. DOE grant No. DE-FG02-91ER40608. Computational cycles were provided by the NERSC high performance computing facility at LBL and by the facilities of the Yale University Faculty of Arts and Sciences High Performance Computing Center.
\end{acknowledgments}

\end{document}